# A Domain Ontology for Modeling the Book of Purification in Islam

Hessa Abdulrahman Alawwad

College of Computer Sience and Information, Imam Mohammad Ibn Saud Islamic University, (IMSIU), Riyadh, Saudi Arabia

*ABSTRACT*

*This paper aims to fill the gap in major Islamic topics by developing an ontology for the Book of Purification in Islam. Many trusted books start with the Book of Purification as it is the key to prayer (Second Pillar after Shahadah, the profession of faith) and required in Islamic duties like performing Umrah and Hajj.*

*The strategy for developing the ontology included six steps: (1) domain identification, (2) knowledge acquisition, (3) conceptualization, (4) classification, (5) integration and implementation, and (6) ontology generation. Examples of the built tables and classifications are included in this paper.*

*Focus in this paper is given to the design and analysis phases where the technical implementing of the proposed ontology is not within this paper's objectives. Though, we presented an initial implementation to illustrate the steps of our strategy.*

*We make sure that this ontology or knowledge representation on the Book of Purification in Islam satisfy reusability, where the main attributes, concepts, and their relationships are defined and encoded. This formal encoding will be available for sharing and reusing.*

## 1. INTRODUCTION

Sharing and exchanging knowledge are two of the basic purposes of communication. People request and give information in daily communication. Comprehending the topics they share and their relationships with other people happens smoothly and naturally with a person's knowledge-building process. Teaching is one of the best-known examples of knowledge exchange.

With the increased use of the internet, people have carried their communicative natures to the digital community, where question-answering systems are one method of revealing their need for information-sharing, and where people, or an intelligent agent that acts on their behalf, share and request knowledge from one or many sources.

As information is distributed by many sources, it is hard for machines to understand it and, therefore, to give the right answer to a specific question. To make the process of understanding this data or information possible, we need to model domain knowledge that comprises the main concept described with attributes and relationships with other concepts.

The Semantic Web was originated by the creator of the World Wide Web, Tim Berners-Lee, who described it as the shift of the web from its conventional, hyperlinked documents, to a huge semantic information storage system where machines can understand and retrieve knowledge from it. Machines should conduct automated reasoning using inference rules on the structured data in order to exploit the Semantic Web. In order to achieve that, knowledge representation must be modeled [1].

Modeling concepts along with their relationships in a way that makes it easy for machines to comprehend them introduces the importance of ontology. Ontology is a philosophical term that has been discussed for a long time, and it refers to the subject of existence, defining a being in a manner that answers the questions: How does this being mean to exist? How does the existence of one being differ from that of another? In computer science, ontology is a structured set of concepts that makes sense for information [2].

Ontology in the context of knowledge sharing, as Tom Gruber has described, is a specification of conceptualization [3]. Representing the information in an ontology makes it both understandable for machines and humans and available for share and reuse.

Ontologies provide a framework for extracting conclusions from the structured information [4]. They mainly reduce the conceptual confusion among those who share the information. They can be applied to many areas, for example knowledge-sharing and reuse, artificial intelligence (AI), software design, and the Semantic Web.

Work on expert systems in the 1980s led AI developers to model knowledge in a standardized semantic manner that machines would be able to understand and comprehend. They also wanted this knowledge to be shared and reused. As ontologies offer this paradigm of knowledge representation and sharing, it was an emerging solution to such an environment.

Ontologies are defined using semantic markup languages like the Resource Description Framework (RDF) and the Web Ontology Language (OWL) and are structured as taxonomies. RDF models the information objects as HTTP Uniform Resource Identifiers (URIs), comprise the information found on the web, and start with "http:." The RDF models the information or resources conceptually and describes them as a statement of (subject, predicate, object) form.

The Resource Description Framework Schema (RDF/S) adds, to some extent, a semantic to the RDF description, so it is an enhanced approach of describing resources based on the RDF definition. RDFS describes the resources along with the relationships among them by defining the domain as classes, subclasses, and properties. The OWL offers much more richness in the semantic description of the relationships between concepts like stating the disjointedness between two classes.

The level of expressiveness of the semantic can be enhanced by the use of the OWL, which is built on top of RDF and RDFS, to model complex relationships. Querying the data stored in the RDF representation is possible using an RDF query language called SPARQL. This could be used to test the built ontology against the domain for which it was developed.

Some authors in [5] identified three major scenarios for the use of ontology: first, supporting the communication process between people, where an unambiguous yet informal ontology would be sufficient; second, achieving interoperability in communication between computer systems by translating between the different modeling techniques, paradigms, languages, and software tools where the ontology is used as an interchange format; third, improving both the process and quality of engineering software systems by meeting the reusability, reliability, and maintainability.

## 2. BACKGROUND

Arabic and/or Islamic ontologies have been active fields of academic research. The WordNet Project [6] introduced an ontology for mapping between different languages. In Quranic ontology, Quran is the religious book of Islam and a revelation from Allah. Knowledge representation defines key concepts in Quran, along with the relationships between these concepts. A few ontologies have been developed [7] [8].

A few ontologies also have been developed for Hadith, the second source of Islamic legislation, which is a record of the words, actions, and silent approval of the Prophet Muhammad PBUH. TibbOnto [9] presents the Prophet's medicine in a semantic ontological representation based on an authentic Tibb Al-Nabawi Hadith. As an example of Islamic ontology, some authors in [10] developed an ontology of the business model, diverse roles, distinct concepts, and representation rules of the Islamic banking industry.

Another group of researchers of Hadith science defined a large Arabic Islamic ontology [11] They followed different steps including defining concepts, relationships, functions, axioms, and instances. They defined five criteria to shed light on the importance of the ontology of clarity, consistency, extensibility, minimal encoding deformation, and minimal ontological commitment. Their ontology was created with Portege's editor and the OWL and included 183 concepts and 145 relations.

With Islam-related research, researchers need to carefully choose their main source or reference for ontology development, whether the Quran, Hadith, Qiyas, or consensus. Commentary books (Kutub Al Shuruh) are also reliable sources of Islamic legislation. The problem with Islamic ontology is the lack of coverage of many Islamic topics. One study on Ketab Al-Salaat [12] proposed an ontology for Salaat (the Second Pillar of Islam) on the Protégé tool based on a test-driven ontology-development methodology (TODE). They have developed a prototype application on the JENA semantic web toolkit for the utilization of the proposed ontology. Their ontology (Figure 1) consisted of 113 concepts and 85 properties.

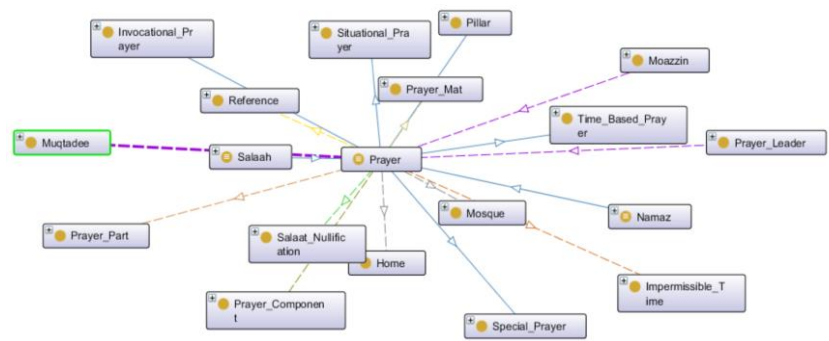

Figure 1. The proposed ontology of Salaat along with properties and relations [12].

Authors in [13] also proposed an ontology for Salaat, their constructing process encompassing three stages: first, determining the domain and scope of the ontology, which are Islamic knowledge and Salaat, or prayer, respectively; second, reusing the existing ontologies where they used Quran indexes; and third, defining the classes, the class hierarchy, and the properties of classes. Finally, they evaluated their method. Their ontology (Figure 2) covered 48 concepts and 51 properties.

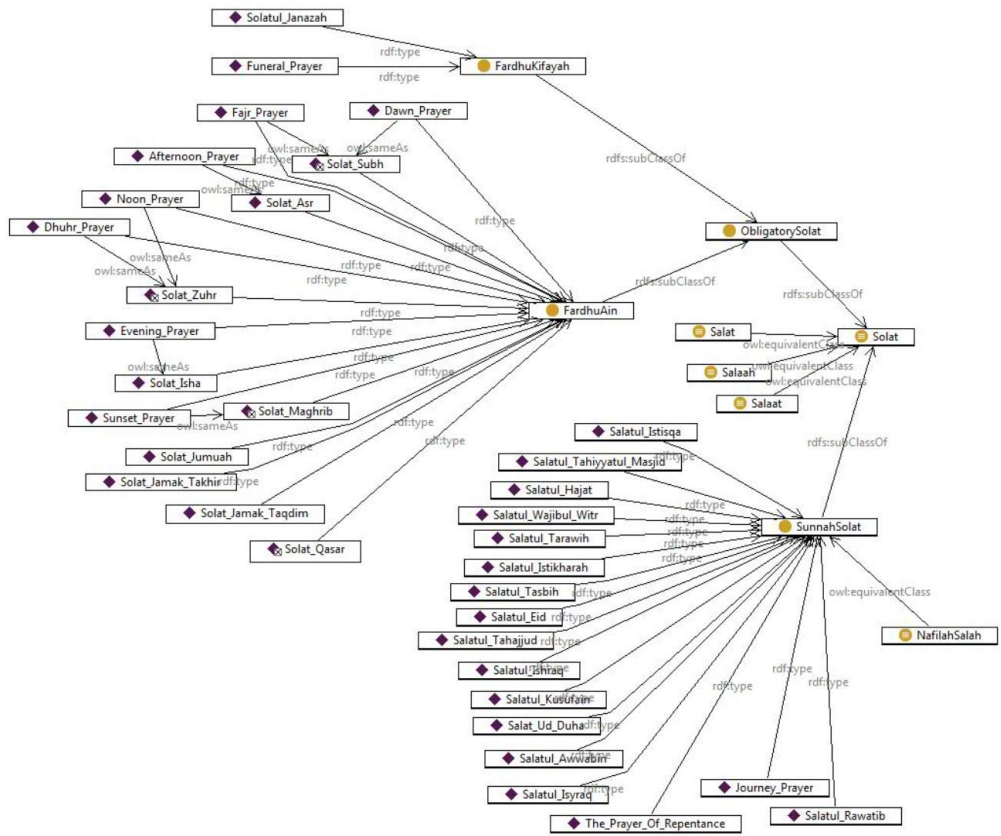

Figure 2. The proposed ontology of Solat along with properties and relations [13].

Authors in [14] defined an ontology for Hadith understanding with a specification word, "Samiaa," which indicates the hearing. The domain of knowledge was Sahih Al-bukhari, which is recognized as the most reliable book of Hadith. They performed five stages of Hadith understanding. In the first phase, they extracted Sahih Al-bukhari from Hadith software. In the second phase, they identified the word Samiaa as the root word and then identified the noun form of Samia and its morphologies. Then, they extracted all Hadith that contained the word Samiaa and analyzed them based on commentary books. Their overall Hadith ontology is shown in Figure 3.

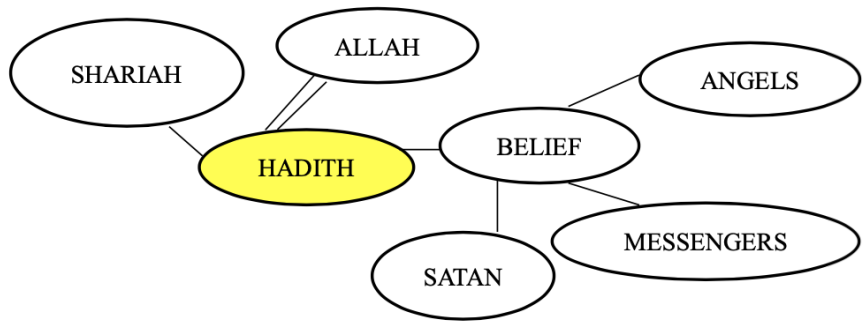

Figure 3. Ontology of Hadith [14].

Other authors [15] proposed an Arabic prophetic ontology about prophets and messengers in Islam based on the Holy Quran, AlHadith, and commentary books. They followed five steps for their ontology building: the first was specification, which means understanding the concepts and their relationship in the Quran, Hadith, and commentary books by meeting experts and reusing related ontologies. Then, they represented the conceptualization in an intermediate representation. Then, they started to build the ontology by defining the class hierarchies and properties of the classes. The resulting ontology contained 151 classes and 44 properties.

## 3. PROBLEM DEFINITION

This paper aims to fill the gap in major Islamic topics by developing an ontology for the Book of Purification in Islam. Many trusted books start with the Book of Purification, as it is the key to prayer (the Second Pillar after Shahadah, the profession of faith) and is required in performing Umrah and Hajj, for example.

## 4. ARCHITECTURE DESCRIPTION

This paper's focus is to ensure that this ontology or knowledge representation on the Book of Purification in Islam has satisfying reusability, where the main attributes, concepts, and their relationships are defined and encoded. This formal encoding will be available for sharing and reuse.

This paper also focuses on meeting the interoperability, reliability, and maintainability attributes. The defined ontology will ease the process of communication between the software systems (interoperability). This formal encoding will also help in the automation of consistency checking, which will make the system more reliable (reliability). The use of the defined ontology in a software system can render maintenance easier in many ways. Building systems by defining an explicit ontology improves documentation of the software, which, in turn, reduces maintenance costs [16].

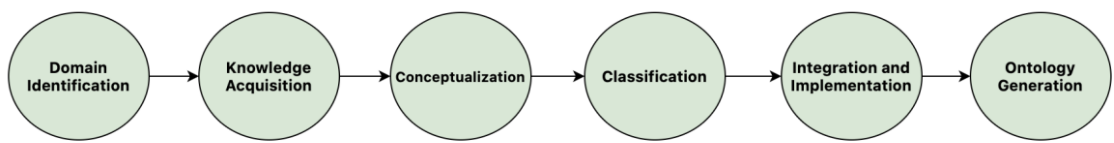

Figure 4. Design and development of the Book of Purification ontology.

Figure 4 presents our methodology for the development of the Book of Purification ontology.

## 4.1. Domain Identification

This first step is concerned with identifying the domain of this ontology and specifying the reference for constructing the ontology. It encompasses the process of defining the domain, scope, and purpose of the ontology and reviewing any existing domain ontologies.

The ontology model for the Book of Purification in Islam provides a semantically structured model that is understandable by both humans and machines for the concepts of purification and facilitates knowledge-sharing and retrieving in the domain of purification.

As this paper aims to develop an ontology for the Book of Purification, the reference that will be used to identify the ontology is the book Alruwd Almurbea. This is one of the Shuruh books written by Imam Mansoor Alhanbali. Other references, like Sahih al-bukhari, Sunan Ibn Majah, and Sunan Abi Dawud, can be included in future work.

## 4.2. Knowledge Acquisition

This step is concerned with acquiring the knowledge needed to build the ontology from the specified reference. This step includes meeting experts and revising knowledge.

## 4.3. Conceptualization

Table 1. Sample of the concept table for the Book of Purification ontology.

| Concept | Description |
|---|---|
| Islam | This is the main class that encompasses all the rest of classes. |
| Pillars | Five basic acts in Islam, considered mandatory by believers, and are the foundation of Muslim life. They are summarized in the famous hadith of Gabriel. |
| Worship | In Islam, worship refers to ritualistic devotion as well as actions done in accordance to Islamic law which is ordained by and pleasing to Allah (God). Worship is included in the Five Pillars of Islam, primarily that of salat, which is the practice of ritual prayer five times daily. |
| Salah | The second of the five pillars in the Islamic faith as daily obligatory standardized prayers. It is a physical, mental, and spiritual act of worship that is observed five times every day at prescribed times. |
| Purification | Cleanliness and being free from dirt. While in the context of Islamic Jurisprudence (Shari'ah), It means the removal of impurities and dirt. |
| Inner Purification | It is the purification of the heart from polytheism, sins etc. |
| Physical Purification | It is the purification of the body from dirt and impurities. |

In this step, we specify conceptualization by defining tables. These tables will be used to directly build the ontology in section 4.5: (1) The concept table (table 1 presents a sample) contains the domain concepts along with the hierarchy of these concepts. Each concept will be defined in a <concept, description> glossary. (2) The properties table (table 2 presents a sample) is used for identifying the internal structure between concepts by specifying the domain and range for every property along with a description for it. (3) A relation table defines constraints, types, and cardinality. (4) Create the individuals of the concepts and their defined relationship.

Table 2. Sample of the properties table for the Book of Purification ontology.

| Object properties | Domain (Concept) | Range (Concept) | Description |
|---|---|---|---|
| achieved_by | Wudu | Purification | Purification from these is achieved by performing al-wudu |
| achieved_by | Ghusl | Purification | Purification from these is achieved by performing Ghusl |
| purify_through | Body | Ghusl | It is possible to purify oneself through wudu or ghusl. |

### 4.4. Classification

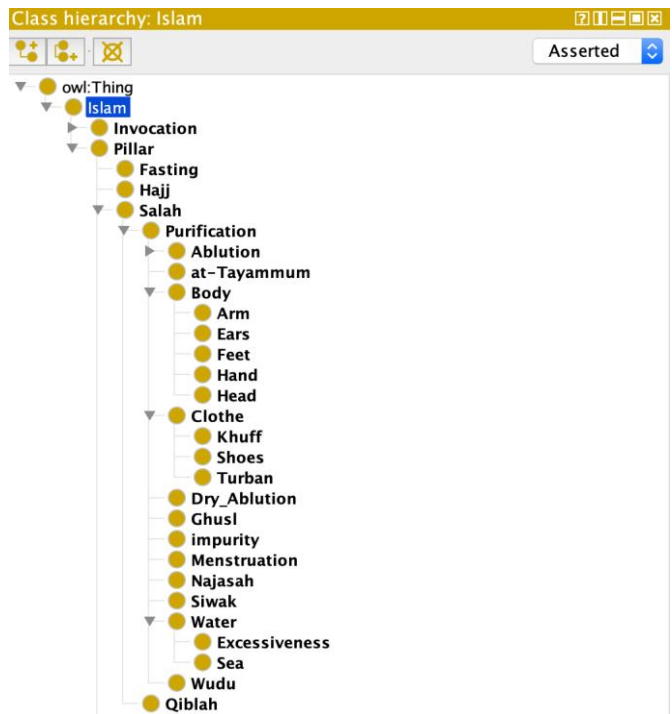

Figure 5. Part of the defined classes along with subclasses.

Classification is the main step of building the ontology, it consists of four steps. (1) Identifying the class hierarchy: the class is the general specification that can be used to instantiate individuals. Figure 5 presents part of the ontology of the Book of Purification and the classes along with subclasses. (2) Defining the properties of classes. Properties represent the relationships in the ontology; this is the second step after defining the classes and subclasses. Property could be either data property or object property. Object property has the type owl:ObjectProperty and links individuals to individuals, whereas data property has the type owl:DatatypeProperty and links individuals to data values. Figure 6 presents part of the defined object and data properties. (3) Creating the individuals of our ontology. Individuals are the instances of our classes. Figure 7 represents individuals for the class Natural_Discharges under the Minor Hadath class. (4) Creating the axioms of our ontology. In order to describe the nature of the relationship between the classes, attribute and instance axioms are used. Axioms are defined for the classes to describe how a certain class relates to another. It could be either a subclass (where a class is a subclass of another class), an equivalent class, or a disjointed class.

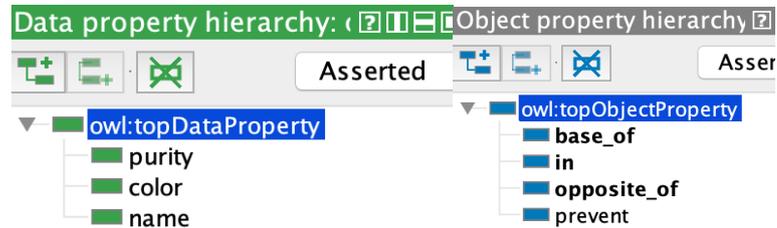

Figure 6. Part of the defined object and data properties.

The class hierarchy in Figure 7 represents some of these axioms, where Minor and Major Hadath are subclasses of the class Hadath.

Axioms for attributes represent how an attribute relates to another, and whether the relationship is transitive, inverse, or equivalent. Axioms for individuals state whether two individuals are the same or different.

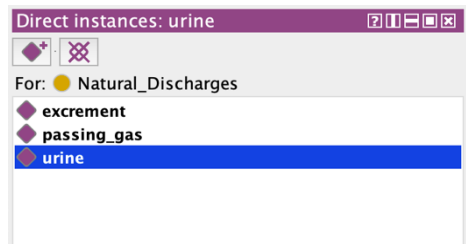

Figure 7. Individuals of the proposed ontology for the Book of Purification.

## 4.5. Integration, Implementation and Ontology Generation

The last two steps include implementing the knowledge defined in the previous steps and possibly re-using an existing ontology like Salaat to expand the applicability of our ontology.

The implementation process will be achieved using the standard ontology language OWL and the ontology editor Protégé, which includes reasoning tools to check the consistency of the built ontology.

Since the technical implementation of the proposed ontology is not within our objectives in this paper, focus is given to the design, analysis, and initial work of the last two steps. The intensive work on Steps 5 and 6 will be left for future work.

## 5. CONCLUSIONS

In this paper, we have discussed the gap that needs to be filled with encoding the major Islamic topics in the formal encoding that is available for share and reuse by means of ontology.

We make sure that the proposed ontology or knowledge representation on the Book of Purification in Islam is defining and encoding the main attributes, concepts, and their relationships.

We have presented in detail our strategy in building the ontology, which encompasses six steps, including domain identification, knowledge acquisition, conceptualization, classification, integration and implementation, and ontology generation.

As designing and analyzing the ontology was the focus of this paper, we started with an initial implementation to represent the steps in our strategy in this paper.

Other systems such as Q&A systems can reuse our ontology. It can be tested and evaluated by user-driven and application-driven approaches.